\documentclass[
    ,final            
  ]
  {aipproc}

\layoutstyle{6x9}

\newcommand{\dslash}[2]{{{#1}\hspace{-5pt}{/}}_{#2}}
\newcommand{\Bsl}{{B\hspace{-5pt}{/}}}
\newcommand{\ndop}{{\mathcal O}_{R/L;L}^{\Bsl}}

\begin{document}

\title{Hadronic matrix elements of proton decay\\ on the lattice}

\author{Yasumichi Aoki\footnote{Present address: Physics Department,
University of Wuppertal, 42097 Wuppertal, Germany}
\hspace{4pt} for RBC collaboration}{
  address={RIKEN BNL Research Center, Brookhaven National Laboratory,
  Upton, NY 11973, USA}
  ,altaddress={Physics Department, Columbia University,
  New York, NY 10027, USA}
}

%

\begin{abstract}
 We report on our on-going project to calculate proton decay matrix
 elements using domain-wall fermions on the lattice.
 By summarizing the history of the proton decay calculation
 on the lattice, we reveal the systematic errors of 
 those calculations. Then we discuss our approach to tackle
 those uncertainties and show our preliminary results on the matrix
 elements.
\end{abstract}

\maketitle



Nucleon decay is one of the most important aspect that 
any (SUSY) GUT model has.
At low energy dimension-six operators are dominant contribution
to the proton decay while higher dimensional operators are suppressed
by the inverse power of the heavy mass ($M_X$).
The dimension-six operators consist of three quark and one lepton
fields. While the lepton part is treated trivially,
the matrix element of the three-quark part
\begin{equation}
 \ndop \equiv \epsilon^{ijk}(u^{iT} CP_{R/L}d^j)P_{L}u^k
\end{equation}
between the initial proton and final pion (K or $\eta$ meson) states
receives a highly non-perturbative contribution from QCD, 
which we want to tackle in this study. 
The matrix element has a tensor structure \cite{Aoki:1999tw},
\begin{equation}
 \langle \pi;\vec{p} | \ndop | p;\vec{k}\rangle = 
  P_{L} [ W_0 - i \dslash{q}{} W_q] u_p,
  \label{eq:ff}
\end{equation}
where $q=k-p$ is the momentum transfer, $u_p$ is the proton spinor.
The relevant form factor
$W_0$ is what we need since the $\dslash{q}{}$ is practically zero 
by the on-shell condition of the outgoing lepton.

The lattice gauge theory gives the first principle computational
ground for the hadronic quantities like this matrix element
of proton decay.  In the first two calculations on the lattice
\cite{Hara:1986hk,Bowler:1988us}, the tree level chiral perturbation
theory \cite{Claudson:1982gh} was used to evaluate $W_0$ from the low
energy constant $\alpha$ and $\beta$, 
\begin{equation}
 \alpha P_L u_p  \equiv  \langle 0 | {\mathcal O}_{R;L}^{\dslash{B}{}} |
  p\rangle, \mbox{\hspace{12pt}} 
  \beta P_L u_p \equiv \langle 0 | {\mathcal O}_{L;L}^{\dslash{B}{}} | p\rangle,\end{equation}
which are calculated on the lattice. This method is sometimes called
the indirect method.
Few years ago, JLQCD published on their large simulation of the
nucleon decay matrix element \cite{Aoki:1999tw}, where they employed
both direct and indirect methods. The direct method was first used by 
the authors of ref.~\cite{Gavela:1989cp}. However, the treatment of the
form factors was improper, which led a large
discrepancy in the results from direct and indirect methods. 
Once the direct method is treated properly, 
JLQCD \cite{Aoki:1999tw} found the discrepancy
not so large, yet, $30-40\%$ in most of the cases. Their results of
the matrix elements are 3--5 times larger than those from
a model calculation commonly used, which pushes down the theoretical
estimate of the life time of the proton, and makes much severe constraint
on the GUT models.

The existing calculations are all done with the Wilson fermion 
at a single lattice spacing. The Wilson fermion has an $O(a)$ 
discretization error, where $a$ is the lattice spacing. There
are two sources of error propagating to the matrix elements.
One is the measurement of the matrix element in lattice unit.
The other is the estimate of the lattice scale $a^{-1}$.
Even for the state of the art calculation by JLQCD, the
systematic error of the $a^{-1}$ is as much as $30\%$ \footnote{The
dimensionless quantity, the product of the Sommer scale
\cite{Sommer:1994ce,Guagnelli:1998ud} and the $\rho$
mass  $r_0 m_\rho$ is about $30\%$ off from its continuum limit
\cite{Aoki:2002fd} at the simulation point of JLQCD. Note that
the decay width is proportional to the square of $W_0$ (dimension two),
or $\alpha$ and $\beta$ (dimension three).}.
Of course there should be a scaling violation for $W_0$,
$\alpha$ and $\beta$, too, which could diminish the overall violation
by compensating that from the scale. But it is unknown until it
is studied.  Also the Wilson fermion breaks chiral symmetry
explicitly. Thus, the applicability of the chiral perturbation theory
at a finite lattice spacing is not guaranteed.
One has to take the continuum limit
of quantities of interest.

The second problem is that up to now the operator
renormalization has been done by one-loop (tadpole-improved) perturbation
theory. This should be improved by employing a non-perturbative
technique \footnote{For the recent summary of the non-perturbative
renormalization on the lattice, see \cite{Sommer:2002en}.}. 

Finally the calculations are all done in the quenched approximation,
where all quark
loop effects are neglected. This approximation is
commonly used in the lattice calculation as the unquenched simulation
is much more expensive. One has to check how large
is the effect of quenching by doing the unquenched simulation. 

\vspace{12pt}

We use the domain-wall fermions \cite{Kaplan:1992bt, Shamir:1993zy, 
Furman:1995ky} in our simulation. This fermion discretization 
has almost exact chiral symmetry and exact flavor symmetry.
Hence there practically is no mixing of the operators with different
chiral structure, making the data cleaner. Moreover, 
there is no $O(a)$ discretization error. 
This second point has been demonstrated in 
the simulation results for the hadron spectrum
\cite{Aoki:2002vt}
and the kaon $B$ parameter
\cite{Blum:1997mz,AliKhan:2001wr}.
The chiral symmetry can be further improved dramatically by improving
the gauge action
\cite{AliKhan:2000iv, Aoki:2002vt}. We use DBW2 gauge action which 
reduce the residual chiral symmetry breaking by factor $100$ from
that for the Wilson gauge action at a typical lattice spacing 
\cite{Aoki:2002vt}.

We use the $16^3\times 32$ lattice with $a^{-1}\simeq 1.3$ GeV
\footnote{The more precise description of our simulation is given in 
ref.~\cite{Aoki:2002ji}.}.
The direct method uses the ratio of the three- and two-point functions
\begin{equation}
 R(t)\equiv\frac{\langle J_{\pi}(t_1) \ndop (t)
  \bar{J_{p}}(t_0)\rangle}
  {\langle J_{\pi}(t_1) J_{\pi}^{\dagger}(t)\rangle
  \langle J_{p}(t) \bar{J_{p}}(t_0)\rangle} \sqrt{Z_\pi Z_p},
  \label{eq:ratio3pt}
\end{equation}
where the proton and pion interpolating fields are
located at $t_0=6$ and $t_1=24$ respectively.
Momentum $\pm\vec{p}$ with $\vec{p}a=(1,0,0)\pi/8$ or $(1,1,0)\pi/8$ is
injected to the pion and the operator in the  
three point function, as well as in the pion two point function 
in the denominator.
$\sqrt{Z_\pi}$ and $\sqrt{Z_p}$ 
are overlap of $J_{\pi}$ and $J_{p}$ 
to the corresponding pion and proton states, which 
is estimated from the fit of two point functions.

\begin{figure}
 \includegraphics[height=.26\textheight]{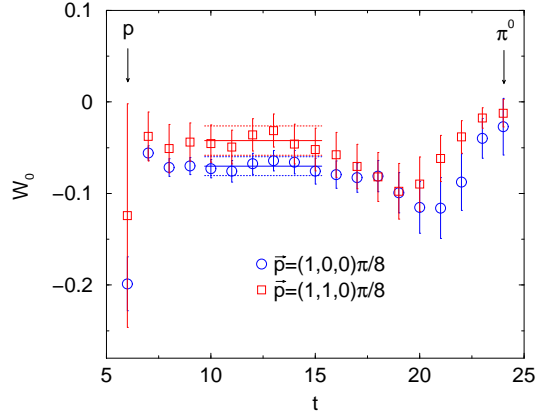}
  \caption{Ratio $R(t)$ for the relevant form factor $W_0$ for $\langle
  \pi^0|{\mathcal O}_{R;L}^{\Bsl}|p\rangle$ at $m_1=m_2=0.04$.}
 \label{fig:p2pi0-t}
\end{figure}

\begin{figure}
  \includegraphics[height=.26\textheight]{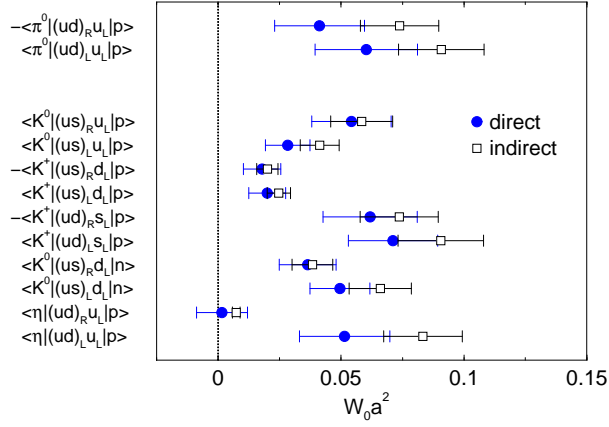}
  \caption{Summary of the relevant form factor of the nucleon decay
 with both direct and indirect methods in lattice unit.}
 \label{fig:summary}
\end{figure}

Figure \ref{fig:p2pi0-t} shows ratio at a parameter point with the
particular projection and subtraction to get $W_0$, 
which is taken from the fit to the plateau.
In addition to the data shown in the ref.~\cite{Aoki:2002ji},
we have further performed the calculation for the non-degenerate quark
mass $m_1$, $m_2$ in the final pseudoscalar state,
where the initial proton state is made up of quarks with $m_1$. 
Then we get $W_0$ as a function of $m_1$, $m_2$, and $q^2$.
The chiral perturbation \cite{Aoki:1999tw} helps to fit $W_0$ to get
to the physical point.
The results for various decay amplitudes are shown in
Fig.~\ref{fig:summary}. We are assuming the SU$(2)$ symmetry for the 
$u$ and $d$ quarks. There are other possible matrix elements, but
they can be calculated with the matrix elements in the figure
when the SU$(2)$ symmetry is intact.
We are yet to have the renormalization factor for the operators by
a non-perturbative renormalization.
Preliminary value using the
perturbative estimate of the renormalization factor \cite{Aoki:2002iq}
is listed in ref.~\cite{Aoki:2002ji}.

The results of the indirect method are also shown in Fig.~\ref{fig:summary}.
The direct and indirect calculations give consistent 
results within the error, in contrast to the result of JLQCD.
However, this could be caused by larger statistical error in our calculation.
We need to have more statistics to judge it.  Nevertheless, the relative
size of the matrix element in our calculation for each decay mode is
similar to that obtained by JLQCD.

\vspace{8pt}

We have investigated the proton decay matrix elements at a 
lattice cut off of $a^{-1}\simeq 1.3$ GeV with the domain-wall
fermion in the quenched approximation. The direct and indirect methods 
give consistent results within our 
precision. The non-perturbative renormalization program 
\cite{Martinelli:1995ty,Blum:2001sr} is underway
to get the continuum matrix elements. 
Also we are performing the two flavor dynamical domain-wall fermion
simulation, which will give us an idea of the size
of the quenching error.


\vspace{8pt}
We thank RIKEN, Brookhaven National Laboratory and the U.S.\
Department of Energy for providing the facilities essential for the
completion of this work.


\vspace{-12pt}
%


\end{document}